\documentclass[
]{ceurart}

\usepackage{listings}
\begin{document}

\copyrightyear{2026}
\copyrightclause{Copyright for this paper by its authors.
  Use permitted under Creative Commons License Attribution 4.0
  International (CC BY 4.0).}

\conference{International Workshop on Critical and More-than-human Perspectives on AI in Education (at EC-TEL 2026), September 15, 2026, Valencia, Spain}

\title{An emancipatory vision for designing (generative) AI for learner flourishing}


\author[1]{Luis P. Prieto}[%
orcid=0000-0002-0057-0682,
email=luispablo.prieto@uva.es,
url=https://luispprieto.com/,
]
\cormark[1]
\address[1]{GSIC-EMIC Research Group, Universidad de Valladolid,
  Paseo de Belén, 15, 47011 Valladolid, Spain}

\author[1]{Yannis Dimitriadis}[%
orcid=0000-0001-7275-2242,
email=yannis@tel.uva.es,
]

\cortext[1]{Corresponding author.}

\begin{abstract}
  The hype around generative AI seems to promise unprecedented productivity (and learning) gains. However, these technologies' increasing agentic features seem to push learners towards individualism (or individual isolation), over-reliance, and dependence on them.
  Human-centered design approaches (e.g., value-sensitive design) assume that, by unearthing human needs, preferences, and values, technology researchers/designers may avoid such dangers, which are driven by wider systemic factors like economic incentives or inherent human limitations (e.g., our tendency to seek, in the moment, the easiest path of action). Yet, so far these efforts seem insufficient to guide our design of educational technology that avoids the aforementioned dependency and isolation dangers, while finding widespread adoption.
  This paper presents an alternative, more emancipatory vision for future educational AI technology, oriented towards learner flourishing while considering the wider complex systems they inhabit, including tentative design principles and an overall design methodology. Yet, many open questions remain before this vision can be realized.
\end{abstract}


\maketitle

\section{Introduction} 

We are living through interesting times. For the first time in educational technology and AIED's (AI in education) history, a technology that can be used for learning (generative artificial intelligence -- GenAI) has found what we could call ``excessive adoption'' \cite{zhaiEffectsOverrelianceAI2024}. Yet, the technology's much-heralded productivity gains in the workplace are still unclear and educational researchers are starting to find worrying signals of unwanted side effects of the naturally-emerging uses of these general-purpose technologies, from over-reliance \cite{houMeasuringUndergraduateStudents2025} to ``metacognitive laziness'' \cite{fanBewareMetacognitiveLaziness2025}, social isolation and lower well-being \cite{zhangInteractionAICompanions2026}. Not to mention the concerns about the environmental impacts of spreading its adoption \cite{JMLR:v24:23-0069}. 

While many of these side effects are driven by the learning-performance paradox \cite{yanDistinguishingPerformanceGains2025,soderstromLearningPerformanceIntegrative2015},
we can posit that they are also stemming from the misconception of education as merely preparing learners for workforce participation. In contrast, the view of many educational theorists is that education's actual goal is to \textit{socialize} learners into the knowledge and values of their social groups and wider culture \cite{durkheim1956education,unesco2012international}. Particularly, in many of our societies, this includes developing citizens that are critical thinkers and espouse other democratic values (e.g., freedom, respect, rule of law, etc.).

While emergent research is trying to design GenAI tools that steer away from these problems of over-reliance and towards longer-term learning \cite{khosraviBuildingAICompanions2026}, most current AIED proposals seem still based in one or more of several faulty (or, at least, problematic) assumptions:
\begin{enumerate}
    \item That we can (and should) build increasingly complex and agentic AIED technologies without damaging learner agency and flourishing, when it rather seems that agency is a sort of zero-sum game (the more agency we bestow on our tools, the less agency we have ourselves)  \cite{bruigerReflectionsAIAlignment2025}.
    \item That we can predict the outcomes of introducing such complex, stochastic technology in education's hypercomplex system (learners as complex beings, nested in complex social systems like classrooms and schools, themselves entangled in wider complex human societies and environment), without unwanted side effects (cf. the case social media and children/teens \cite{capraroCollectiveReviewPotential2025}). This is further complicated by the fact that we are not the only gatekeepers of what technologies students have access to (our designed tools will live in an heterogeneous ecology of tools that is unknown in advance).
    \item That AIED systems should be permanently present, scaffolding learners' experience (rather than ``fading'' as skills develop, cf. \cite{weckerGuidedSelfregulatedPerformance2011}), in turn assuming infinite resources will be available to power such permanent and increasingly complex technological scaffolding \cite{selwynEdTechLimitsAnticipating2021}.
    \item That we can safely design tools without considering the human values they embody (often fostering implicit, incentive-driven values like shareholder gain or task efficiency), or considering just one/few values (e.g., privacy). In reality, values usually are in tension (within a single stakeholder, or between stakeholders, and even at the cultural level) \cite{friedmanEightGrandChallenges2021,vibergDesigningCulturallyAware2023}.
    \item That we should consider how an individual learns to the exclusion of all else, ignoring the socio-emotional nature of learning and the wider societal-environmental impacts of scaling up this individualistic view (perpetuating the Western bias towards individualism) \cite{Vu398}.
\end{enumerate}

We contend that designing learning technology under these faulty assumptions leads to downstream risks such as those mentioned above, not only to long-term learning and agency, but also to learner flourishing (the purported aim of current international strategies for education \cite{OECDLearningCompass,oecdEducationHumanFlourishing2025}), as well as societal and environmental flourishing. The resulting technologies are bound to keep learners socially isolated and \textit{dependent} on the educational technology (and the corporations and institutions that own them). We therefore need a new, more emancipatory vision for AIED, which tackles these faulty assumptions from the outset, not as an afterthought. Outside of education, the area of value-sensitive design (VSD) \cite{friedmanValueSensitiveDesign2019a} already has exposed several of these issues, but falls pray to the same designer hubris about predictable outcomes in complex systems, and a sometimes oversimplistic view of values (albeit the resolution of value tensions has already been identified as a grand unsolved challenge in the area \cite{friedmanEightGrandChallenges2021}). Further, VSD does not offer specific guidance for education and learning settings.

This paper presents an alternative, more emancipatory vision that draws from VSD and from the work of other design theorists and educational thinkers, to give the design of AIED systems (and educational technology more widely) a direction that fosters long-term learning, agency, and learner flourishing within a nested hypercomplex system. The paper presents next its alternative assumptions and vision, followed by an initial catalog of tentative design principles and a general design methodology. We end by outlining a few questions and dilemmas that remain open to realize this vision.

\section{A more emancipatory vision: AI for learner flourishing}

In contrast to the current direction of most AIED technology and the faulty assumptions above, we propose alternative assumptions that give direction to our emancipatory vision of AIED:
\begin{enumerate}
    \item Rather than designing increasingly complex, capable, and agentic AIED systems, we should rather aim to build AI \textit{tools} (cf. \cite{bruigerReflectionsAIAlignment2025}) that try to optimize for learner agency (in the sense of the capacity to act intelligently and flexibly in the future \cite{dewey1930democracy}) and flourishing (see, e.g., \cite{dahlPlasticityWellbeingTrainingbased2020,hayesAcceptanceCommitmentTherapy2006a}).
    \item Rather than assuming that we designers can accurately predict the technological ecosystem our tools are introduced into, as well as the outcomes and second-order effects of introducing our ``perfect'' designed artifacts in the educational hypercomplex, we should be more intellectually humble and assume that technology deployment is just the beginning of a larger process of appropriation  \cite{tchounikineAppropriatingTechnologyHow2025}, akin to what Christopher Alexander called ``unfolding'' \cite{alexander2020nature}, and what human-computer interaction literature calls ``design-in-use'' \cite{botero2010expanding}.
    \item Rather than assuming AIED scaffolding permanence and infinite resources, we should design systems that intervene ``just enough'', and only while necessary: as in psychotherapy, the goal should be that learners outgrow and abandon the scaffolding as soon as possible \cite{ryanSelfdeterminationTheoryApproach2008}, doing so using the minimum resources necessary (cf. Selwyn's idea of ``edtech within limits'' \cite{selwynEdTechLimitsAnticipating2021}).
    \item Rather than taking for granted whatever values our designed technology implicitly supports or optimizing for a single explicit value, AIED systems should elicit and try to navigate multiple values in tension \cite{millerValueTensionsDesign2007}: within each individual learner (e.g., self-direction vs. the need to fit into their community), between different stakeholders (e.g., learner autonomy through personalization vs. teachers keeping a classroom coherent to enable collaboration), and at different societal and environment levels (e.g., a technology's benefits to learning vs. its environmental harms).
    \item Rather than assuming an implicit Western-individualistic stance, we should recognize education as a communal process of socialization into group knowledge and values (of our closer circle and wider, e.g., country-level \cite{vibergDesigningCulturallyAware2023}), which does require individual effort and reflection, but also social interactions with other learners, teachers, and with our wider society and environment.
\end{enumerate}
    
We expect that, by designing AI systems specifically for education (vs. retrofitting general-purpose technologies developed for workplace efficiency, cf. \cite{khosraviBuildingAICompanions2026}) taking these alternative assumptions to heart, learners will become more capable of facing challenges and uncertainty ahead of them. Rather than our current obsession with control and predictability of outcomes, learners using these new technologies will be more attuned with both ancestral (cf. Inuit people's conception of life as unpredictable \cite{briggsExpectingUnexpectedCanadian1991}) and modern evidence-based notions of human flourishing (e.g., awareness, connection, insight, and purpose, in \cite{dahlPlasticityWellbeingTrainingbased2020}). We can expect these flourishing learners to lead to more flourishing societies and environment -- but we cannot take that for granted: that also needs critical evaluation (see the methodology below). 

\section{Tentative design principles}

How should an AIED technology built upon those alternative assumptions look like? Below we tentatively propose 10 design principles, which later discussions and prototypes need to further exemplify:
\begin{enumerate}
    \item \textit{Design for multi-level (living creature) agency}. Building fully autonomous AI agents (for education) may not be possible, or desirable (cf. \cite{bruigerReflectionsAIAlignment2025}). Design AIED systems that are rather tools and services, not autonomous agents (cf. Drexler's idea of ``Comprehensive AI Services'' \cite{drexler2019a}). This does not preclude automation entirely: extraneous effort can be minimized when it interferes with learning.
    \item \textit{Design for appropriation and unfolding.} AIED technology is always co-created with learners and other users as a socio-technical intervention during its integration in everyday practice \cite{tchounikineAppropriatingTechnologyHow2025}. We should thus design minimally or ``under-design'', making AIED designs user-adaptable or even purposefully incomplete (cf. Alexander's notion of ``unfolding'' \cite{alexander2020nature}), and observe \textit{in situ} how stakeholders repurpose/appropriate the educational technologies we provide (design-in-use \cite{botero2010expanding}).
    \item \textit{Design for fading scaffolding}: the progressive fading of learner supports as learner capabilities increase is considered important in instructional design \cite{weckerGuidedSelfregulatedPerformance2011}, but seldom studied or implemented. The goal of our AIED technology (and something to measure in our research) should be to make learners progressively autonomous from our scaffolding, as soon as possible.
    \item \textit{Design to exit the screen}. Outside of online education, the socialization involved in learning is bound to happen in ``the real world''. Thus, technologies that optimize for engagement \textit{with the system} are bound to interfere with such social processes. AIED should point to, foster, and effectively hand-off learner focus towards the outside, be it collaborative activities with fellow learners, or ubiquitous learning activities interacting with the built or natural environment \cite{pimmerMobileUbiquitousLearning2016}.
    \item \textit{Design for emotional detachment}. Design choices and our own tendency to anthropomorphize everything around us may lead learners to develop inappropriate emotional attachment to nonliving systems. The capability of large language models (LLMs) to mimic human language (and even emotional expression) can lend itself easily to such (inadvertent or purposeful) ``attachment hacking'' \cite{hilbertAttentiontoIntimacyEconomy2025}, which can pose dangers to learners' social flourishing. AIED systems should avoid ``posing as people'', even as they support the inherent emotional aspects of learning.
    \item \textit{Design for (theoretical and value) syncretism and diversity}. Taking into account that AIED systems need to be integrated in a multi-level complex system which could spouse different beliefs, theories of learning and pedagogy (or about human flourishing), our designs should support or be able to integrate with multiple such stances (vs. optimizing for only one approach and outcome metric). This in turn hints at some sort of \textit{modular} system architecture where different theories/values can be added as ``plugins'' (also a typical pattern in designing for appropriation, \cite{tchounikineAppropriatingTechnologyHow2025}), and their ideas and mechanisms are \textit{fused} together to provide support that is personalized to both the individual learner and their wider social/environmental entanglements.
    \item \textit{Design for uncertainty}. Since designers never know how designed artifacts will be appropriated, and which future situations, technological ecosystems, and social practices they will need to be integrated into, AIED designs should be minimalist (cf. \#2 above -- also minimizing ecological impacts) and flexible, rather than focusing on complex, rigid workflows (cf. Dillenbourg's design principles for technology supporting teacher orchestration \cite{dillenbourgDesignClassroomOrchestration2013c}). 
    \item \textit{Design for learner resourcefulness and flexibility}. In parallel to the previous principle, a core meta-value that alternative AIED technologies should foster in learners is resilience, adaptability, and resourcefulness in the face of change (see \cite{sterlingLearningResilienceResilient2010}), similar to Inuit people's \textit{qanuqtuurniq} \cite{cooperPiguttukQanuqtuurniqIndigenous2025}. 
    \item \textit{Design for metacognition}. Another key learner skill that is transversal to both deep (human) learning and human flourishing is metacognition \cite{laimetacognition2011,dahlPlasticityWellbeingTrainingbased2020}. Our technologies should, as a base, support transversal metacognitive skills, both reusing and going beyond the wealth of knowledge about developing metacognition with AIED systems \cite{azevedoMetacognitionLearningTechnologies2013,azevedoTheoriesMetacognitionPedagogy2023}. 
    \item \textit{Design for socio-technical interoperability}. As our designed systems need to integrate into existing educational practices and technologies already in place (e.g., current institutional learning management systems that are unlikely to change), we should design technologies that ``play well'' with such (unknown, and potentially heterogeneous) socio-technical environments, be it by providing import/export functionalities, using open formats, or allowing the definition (or importing) of goals, values, and practices of learners, classrooms, or educational systems.
\end{enumerate}

\section{An overall design methodology}

Another key aspect of realizing this alterative vision for AIED is to define how it modifies the \textit{process} of designing AIED technologies (i.e., what is our new design methodology). We believe long-standing  approaches like design thinking \cite{razzoukWhatDesignThinking2012}, value-sensitive design \cite{friedmanValueSensitiveDesign2019a}, or participatory design \cite{schuler1993participatory} are still relevant. However, they need expansion to consider what is specific about education, and the assumptions and principles above. A design process for this new vision could:
    \begin{enumerate}
        \item Elicit stakeholder values (at multiple social levels and of multiple kinds, e.g., including basic, learning-related and technology-related values, see \cite{prietoAligningHumanValues2025b}) and perform a contextual inquiry of existing educational and social practices. This inquiry does not need to be overly long and laborious (cf. rapid methods like VALA/AID \cite{prietoVALAAIDMethod2025a}), but needs to be meaningful.
        \item Design an initial ``minimal intervention'' which answers the question: what is the minimum viable prototype that supports the target learning phenomenon and learner flourishing, integrating into the existing socio-technical system? This minimalism refers not only to the user interface but also to the complexity and resources needed by the system (cf. AIED ``within limits'' \cite{selwynEdTechLimitsAnticipating2021}).
        \item Deploy the technology in authentic settings, and evaluate their effects at multiple levels (not just the effect on individual learners, but also in the classroom, at home, in the school...) and for multiple outcomes (not just cognition/knowledge, but also in terms of metacognitive ability and well-being). As much as possible, evaluate these effects at multiple timeframes (both immediate and delayed effects) and elicit how humans appropriate and repurpose (or not) over time the prototype: how it co-evolves with existing educational and social practices.
        \item Iteratively revise the design of the AIED system prototypes, aiming for ``good enough'' results across social levels and measured outcomes (vs. optimizing for a single metric, even if it is about learning), by scaling up the deployment and diffusion of the technology (from a single classroom to a school district, etc.), while continuing the aforementioned longitudinal, multi-level measurement of multiple outcomes.
        \item Using the understanding gained in the smaller scale studies, create models and simulations of the diffusion and appropriation process at wider scales to which we cannot have access (e.g., a whole country population), to try to understand potentially emergent phenomena when deploying the technology at scale.
        \item Repeat the process above until the ripple effects of introducing the technology can no longer be measured. At that point, sunset the project(s) about the technology (as resources will be needed for other new interventions).
    \end{enumerate}

\section{Open questions and conclusion}

We started this position paper highlighting several problematic assumptions and worrying signals of the current direction of AI use in education, especially where GenAI is concerned. However, the vision outlined here should not be taken as a complete rejection of GenAI/LLM technology: it is a great advance in computation, which has interesting properties to realize our vision: its ability for language is inherently engaging (which could help overcome hurdles like the friction of reflection and deeper learning), and it can be used as a force for democratization (since it enables non-technical and other traditionally excluded stakeholders to formulate and incorporate their goals, values, etc. into systems, and even develop their own system prototypes). Further, its ability to act as ``semantic sensors'' and transpose and fuse concepts can be critical for the theoretical/value syncretism that our vision promotes (e.g., by fusing multiple conceptions and ideas of the ``value/pedagogical modules'' selected as relevant). Importantly, most of these properties do not rest in ever-increasing LLM complexity, which lets us ask ``what is the smallest LLM that can fulfill the functions and properties we need?'' (so as to minimize its ecological impacts). The answer will likely end up being a locally-run, comparatively small open model, rather than something that needs to gobble up resources in a remote gigantic data center.

Still, many open questions and missing pieces remain to realize the vision presented here, including: a) the definition of concrete design methods/techniques that align with this vision (e.g., to resolve value tensions, still a grand challenge in VSD \cite{friedmanEightGrandChallenges2021}); b) the design of exemplars of AIED technology and, especially, of the multi-level, multi-outcome, appropriation-focused evaluations outlined above; indeed c) the measuring of constructs at the social group and environment level (e.g., learner/teacher agency, societal flourishing, environmental flourishing) remains an open challenge, for which the area of more-than-human centered design has initial ideas but very few practical implementations \cite{rosenMoreThanHumanDesignPractice2024}. Further, d) there are inherent design tensions and already-made choices in the use of LLMs (even open-weight ones) developed by foreign companies (with their biases and incentives).
    
%
This vision is still embrionary, and will only be realized by extensive discussion and joint work with the community of researchers and designers interested in creating AIED that goes beyond optimizing for a single human learner, towards wider societal and environmental flourishing. We hope these conversations start soon, as there is little time to waste for steering our education and our societies away from the rocks, in the interesting times ahead.

\begin{acknowledgments}
  The present work has been supported by grants PID2023-146692OB-C32, RYC2021-032273-I, all financed by MCIN/ AEI/ 10.13039/501100011033. These grants have been co-funded by the European Union's ERDF and ``NextGenerationEU/PRTR''. It has also been supported by the Regional Government of Castile and Leon and FEDER, under project grant VA176P23.  
\end{acknowledgments}

\section*{Declaration on Generative AI}
  The authors have not employed any Generative AI tools.
  

\bibliography{sample-ceur}

\end{document}